\documentclass[
  reprint,  
 amsmath,amssymb,
 aps,
 ]{revtex4-1}

\usepackage{dcolumn}
\usepackage{color}
\usepackage{bm}
\usepackage{xspace}

\usepackage{graphicx}           % for arxiv
\usepackage[hang,small,bf]{caption}
\usepackage[subrefformat=parens]{subcaption}
\newcommand{\be}{\begin{equation}}
\newcommand{\ee}{\end{equation}}
\newcommand{\bea}{\begin{eqnarray}}
\newcommand{\eea}{\end{eqnarray}}
\newcommand{\beaa}{\begin{eqnarray*}}
\newcommand{\eeaa}{\end{eqnarray*}}
\newcommand{\alphaI}{{$\alpha$-(BEDT-TTF)$_2$I$_3$}\xspace}

\begin{document}

\title{Theory for Planar Hall Effect in Organic Dirac Fermion System}

\author{Yuki Nakamura}
\email{nakamura.yuki.84v@st.kyoto-u.ac.jp}
\author{Takao Morinari}
\email{morinari.takao.5s@kyoto-u.ac.jp}
\affiliation{
  Course of Studies on Materials Science, 
  Graduate School of Human and Environmental Studies, 
  Kyoto University, Kyoto 606-8501, Japan
}

\date{\today}

\begin{abstract}
  In a recent experiment on the interlayer magnetoresistance
  in the quasi-two-dimensional organic salt, \alphaI,
  it has been observed that at low temperatures,
  interlayer tunneling attains phase coherence,
  leading to the emergence of a three-dimensional electronic structure.
  Theoretically and experimentally it has been suggested that
  the system exhibits characteristics of a three-dimensional Dirac semimetal
  as a consequence of broken time-reversal symmetry and inversion symmetry.
  Here, we perform a theoretical calculation of the magnetoconductivity
  under an in-plane magnetic field and demonstrate that the system displays
  a planar Hall effect.
  Our calculations are based on a realistic model for \alphaI
  incorporating interlayer tunneling
  and the tilt of the Dirac cone.
  Given that the planar Hall effect is anticipated as a consequence of chiral anomaly,
  our findings provide support for the classification of \alphaI as
  a three-dimensional Dirac semimetal.
\end{abstract}

\maketitle

Massless Dirac and Weyl semimetals have been extensively studied recently
because of their unique and
intriguing electrical properties
\cite{Murakami2007,Burkov2011,Yang2011,Xu2011,Young2012a,Wang2012,Wang2013,Neupane2012,Armitage2018}.
The energy spectrum in these systems is characterized by
the touching of the valence band and conduction band at discrete momentum points.
The key distinction between the Dirac/Weyl semimetal and the two-dimensional
Dirac fermion system lies in the presence of broken time-reversal symmetry
and/or inversion symmetry.
To realize a Weyl semimetal, it is necessary to break 
either time-reversal symmetry or inversion symmetry, or both. 
On the other hand, a Dirac semimetal can be realized 
even when both time-reversal and inversion symmetries are preserved.

Organic charge-transfer salt, \alphaI, has been studied
as a quasi-two-dimensional Dirac fermion system\cite{Katayama2006,Kobayashi2007,Kajita2014}.
(Here, BEDT-TTF is bis(ethylenedithio)tetrathiafulvalene.)
One of the present authors theoretically predicted\cite{Morinari2020,Morinari2023flux} that
both time-reversal symmetry and inversion symmetry are broken,
and, as a result, the system becomes a three-dimensional Dirac semimetal
when the interlayer tunneling becomes phase coherent at low temperatures.
The phase coherence in the interlayer tunneling is confirmed experimentally\cite{Tajima2023}
by the observation of the peak structure in the interlayer magnetoresistance.
Furthermore, the observation of
the negative magnetoresistance and the planar Hall effect (PHE)
has been reported recently\cite{Tajima2023a}
that is associated with chiral anomaly\cite{Nielsen1983,Son2013,Burkov2015,Nandy2017,Burkov2017,Xiong2015,Huang2015a,Hirschberger2016,Zhang2016,Li2016} in a Dirac semimetal.

In this Letter, we consider a model that includes inter-layer tunneling and
the tilt of the Dirac cone that exists in \alphaI\cite{Katayama2006,Kobayashi2007}.
Based on the semiclassical Boltzmann equation, we compute the magnetoconductivity
under in-plane magnetic fields.
We show that the system exhibits a PHE
using a set of realistic parameters for \alphaI.

In the absence of the interlayer tunneling,
there are two Driac cones in the $k_x$-$k_y$ plane\cite{Katayama2006,Kobayashi2007}.
  Upon incorporating interlayer tunneling between both the same and different molecules,
  four Dirac cones emerge, as detailed below.
In contrast to systems where spin degeneracy is lifted due
to the breaking of time-reversal symmetry caused by magnetic correlations,
the spin remains degenerate in \alphaI because the time-reversal symmetry breaking
is not associated with magnetic correlations\cite{Morinari2020}.
For the sake of simplicity, we neglect the spin degrees of freedom in the follwoing analysis.

The Hamiltonian for two of the four Dirac cones is given by
\begin{align}
H(\bf{k})&=\hbar vk_x\tau_x+\hbar vk_y\tau_y-2t_2\cos(ck_z)\tau_z\notag\\
&\quad+\left[-2t_1\cos(ck_z)+\hbar uk_x\right]\tau_0 + \varepsilon_{\rm D}.
\label{Eq:Hamiltonian}
\end{align}
Here $k_x$ and $k_y$ are in-plane wave numbers measured from the Dirac point
and $k_z$ is the wave number perpendicular to the $k_x$-$k_y$ plane.
We note that the position of the Dirac point in the plane is irrelevant
for the following calculation,
though we need to include them to make clear the presence of
the symmetry breaking.
The parameter $u$ describes the tilt of the Dirac cone to the $k_x$ axis,
and we neglect anisotropy in the Dirac cone in the plane.
$c$ is the lattice constant in the $c$-axis.
$\tau_x,\tau_y,\tau_z$ are the Pauli matrices and $\tau_0$ is
the $2 \times 2$ identity matrix.
$t_1$ and $t_2$ are the parameters for the interlayer tunneling.
$t_1$ is for the tunneling between the same molecules,
and $t_2$ is for the tunneling between the adjacent molecules along the $a$-axis.
  When $t_1 \neq 0$ and $t_2 = 0$, the Dirac points shift along lines
  that are parallel to the $k_z$ axis\cite{Kobayashi2008}. 
  If $t_2 \neq 0$, the Dirac fermions acquire mass, with the exception
  at points where $k_z = \pm \pi/2$.
  Consequently, four Dirac points emerge within the three-dimensional Brillouin zone.
%----------------------------------------------------------------------
The Dirac cone is type-I in the $k_x$-$k_y$ plane\cite{Tajima2018}, so 
the range of the parameter $u$ is $-v<u <v$.
The other two Dirac cones are described by Eq.~(\ref{Eq:Hamiltonian})
with $k_x \rightarrow -k_x$.
We may assume $t_1 > t_2$ from the crystal structure of \alphaI\cite{Bender1984}.
In this case, the Dirac cone is type-II\cite{Soluyanov2015} in the $k_z$ direction.
The parameter $\varepsilon_{\rm D}$ denotes the energy of the Dirac point.
We assign different values of $\varepsilon_{\rm D}$ to the two Dirac cones
in the $k_x$-$k_y$ plane to incorporate the symmetry breaking.

The energy dispersion is given by
$E_{\bf{k}}^{\left(\pm\right)} = (\hbar v/a)\tilde{E}_{\bf{k}}^{\left(\pm\right)}$
where
\begin{align}
  \tilde{E}_{\bf{k}}^{\left(\pm\right)}
  =\pm\tilde{E}_{\bf{k}}-2\tilde{t_1}\cos(ck_z)+\eta a k_x + \tilde{\varepsilon}_D,
  \label{Eq:Energy}
\end{align}
with $\tilde{\varepsilon}_D = {\varepsilon}_D/(\hbar v/a)$ and
\be
\tilde{E}_{\bf{k}}=\sqrt{a^2(k_x^2+k_y^2)+4\tilde{t_2}^2\cos^2(ck_z)}.
\ee
Here, $a$ is the in-plane lattice constant.
We take the same lattice constants for $a$ and $b$ axes for simplicity.
We defined the following dimensionless parameters,
\begin{align}
  \tilde{t_1}&=\frac{t_1}{\hbar v/a}, & \tilde{t_2}&=\frac{t_2}{\hbar v/a},
  & \eta&=\frac{u}{v}.
  \label{Eq:Dimensionless}
\end{align}
Taking $a=1.0 \times 10^{-9}$ m and $v=5.0 \times 10^4$ m/s,
we find $\hbar v/a = 3.3\times10^{-2}$ eV.

\begin{figure}[htbp]
  \includegraphics[width=0.5\textwidth]{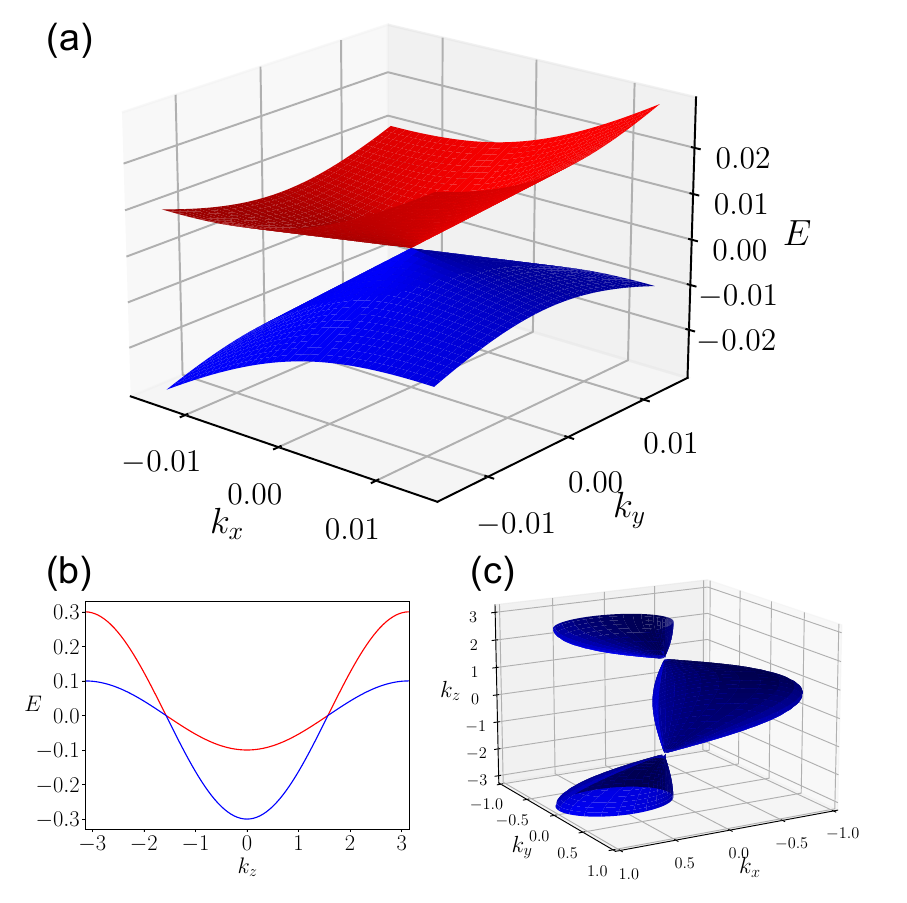}
  \caption{
    \label{fig:EnergyDispersion}
  (Color online) (a) Energy dispersion of the model described
  by the Hamiltonian (\ref{Eq:Hamiltonian})
  in the $k_x$-$k_y$ plane and (b) along the $k_z$ axis.
  The tilt parameter in the plane is $\eta=0.7$, and so
  the Dirac cone is type-I.
  The interlayer hopping parameters
  are $\tilde{t_1}=0.10$ and $\tilde{t_2}=0.05$,
  and so the Dirac cone is type-II in the $k_z$ axis.
  (c) The Fermi surface around the two Dirac cones.
  The Fermi energy is set to be zero
  and we set $\varepsilon_{\rm D} = 0.03$.
  The Fermi surface consists of three portions:
  the middle one is the electron Fermi surface
  and the other two are the hole Fermi surfaces.
  }
\end{figure}

Figure~\ref{fig:EnergyDispersion}(a) shows the energy dispersion in the plane
and Fig.~\ref{fig:EnergyDispersion}(b) shows that in the $k_z$ direction.
We see that the Dirac cone is type-I in the $k_x$-$k_y$ plane
and type-II in the $k_z$ axis
as stated above.
Figure~\ref{fig:EnergyDispersion}(c) shows the Fermi surface.
If the Fermi energy is larger than $t_1$ and $t_2$,
the Fermi surface is a warped cylinder\cite{Tajima2023}.
For \alphaI, the Fermi energy is expected to be smaller than $t_1$ and $t_2$\cite{Tajima2023}.
In this case, the Fermi surface splits into a single electronic Fermi surface
and two hole Fermi surfaces as shown in Fig.~\ref{fig:EnergyDispersion}(c).
Because of the tilt parameter $\eta$, which is slightly lower than one\cite{Tajima2018}, 
the Fermi surface is largely deformed.

%------------------------------------------------------------------------
%------------------------------------------------------------------------

We calculate the magnetoconductivity using the semiclassical Boltzmann equation
employing the relaxation time approximation.
The application of the Boltzmann equation is justified
when $\omega_c \tau < 1$ with $\omega_c$ being the cyclotron frequency
and $\tau$ being the scattering time.
Therefore, our result is limited to the regime of relatively weak magnetic fields.
In the presence of the electric field ${\bf E}$ and the magnetic field ${\bf B}$,
the quasiclassical equation of motion is given by\cite{Duval2006,Xiao2006}
\bea
\hbar \frac{{d{\bf{k}}}}{{dt}}
&=& \frac{1}{{1 + \frac{e}{\hbar }{\bf{B}} \cdot
    \bf{\Omega_k}
}}
\nonumber \\
& & \times \left[ { - e{{\bf{v}}_{\bf{k}}} \times {\bf{B}} - e{\bf{E}} - \frac{{{e^2}}}{\hbar }\left( {{\bf{B}} \cdot {\bf{E}}} \right)
  \bf{\Omega_k}
  } \right],
\label{PHE:BTF:eq1}
\\
\frac{{d{\bf{r}}}}{{dt}}
&=& \frac{1}{{1 + \frac{e}{\hbar }{\bf{B}} \cdot
    \bf{\Omega_k}
}}
\nonumber \\
& & \times
\left[ {{{\bf{v}}_{\bf{k}}} + \frac{e}{\hbar }\left( {
      \bf{\Omega_k}
      \cdot {{\bf{v}}_{\bf{k}}}} \right){\bf{B}} + \frac{e}{\hbar }{\bf{E}} \times
    \bf{\Omega_k}
  } \right],
\label{PHE:BTF:eq2}
\eea
where $\bf{\Omega_{\bf{k}}}$ is the Berry curvature.
%------------------------------------------------------------------------

From the energy dispersion (\ref{Eq:Energy}), the group velocity is given by
\begin{align}
{\bf{v}}_{\bf k}^{\left(\pm\right)}=&v\left(\pm\frac{ak_x}{\tilde{E}_{\bf{k}}}+\eta,\pm\frac{ak_y}{\tilde{E}_{\bf{k}}},\right.\notag\\
&\left.\mp\frac{4\frac{c}{a}\tilde{t_2}^2\sin(ck_z)\cos(ck_z)}{\tilde{E}_{\bf{k}}}+2\frac{c}{a}\tilde{t_1}\sin(ck_z)\right).
\label{Eq:GroupVelocity}
\end{align}
The Berry curvature\cite{Xiao2010} is given by
\begin{align}
  {\bf{\Omega}}_{\bf k}^{\left(\pm\right)}
  =&\left(\mp\frac{2a^2c\tilde{\tau_2}k_x\sin(ck_z)}{2\tilde{E}_{\bf{k}}^3},\mp\frac{2a^2c\tilde{\tau_2}k_y\sin(ck_z)}{2\tilde{E}_{\bf{k}}^3},\right.\notag\\
&\quad\left.\pm\frac{2a^2\tilde{\tau_2}\cos(ck_z)}{2\tilde{E}_{\bf{k}}^3}\right)\label{Eq:BerryCuvture}.
\end{align}
Here, ${\bf{v}}_{\bf k}^{\left(+ \right)}$
%and $\bf{\Omega_k}^{\left(+ \right)}$
and ${\bf{\Omega}}_{\bf k}^{\left(+ \right)}$
are for the positive energy state, $\tilde{E}_{\bf{k}}^{\left( + \right)}$,
and ${\bf{v}}_{\bf k}^{\left(- \right)}$
and ${\bf{\Omega}}_{\bf k}^{\left(- \right)}$
are for the negative energy state,
$\tilde{E}_{\bf{k}}^{\left(- \right)}$.

%------------------------------------------------------------------------
  Now we consider the contribution from the chiral anomaly 
and omit the term related to the anomalous Hall effect. 
From the Boltzmann equation, we obtain the equations 
for the magnetoconductivities\cite{Nandy2017,Burkov2017}:
%------------------------------------------------------------------------
\bea
\sigma_{xx}^{(\pm)}
&=& \frac{2e^2\tau}{(2\pi)^3}
\int d^3{\bf{k}}
%\left( -\frac{\partial f_{\rm{eq}}}{\partial\epsilon} \right)
\left[ - f^{\prime}_{\rm{eq}} \left( E_{\bf{k}}^{\left(\pm\right)} \right) \right]
\frac{1}{1+\frac{e}{\hbar}{\bf{B}}\cdot
  {\bf{\Omega}}_{\bf k}^{\left(\pm\right)}} \nonumber \\
& & \quad\times
\left[
  v_x^{\left(\pm\right)}
  +\frac{e}{\hbar}B_x(
  {\bf{v}}_{\bf k}^{\left(\pm\right)} 
  \cdot{\bf{\Omega}}_{\bf k}^{\left(\pm\right)} )
  \right]^2,
  \label{eq:sxx}
\eea
\bea
\sigma_{xy}^{(\pm)}
&=&\frac{2e^2\tau}{(2\pi)^3}\int d^3{\bf{k}}
%\left(- \frac{\partial f_{\rm{eq}}}{\partial\epsilon}\right)
\left[ - f^{\prime}_{\rm{eq}} \left( E_{\bf{k}}^{\left(\pm\right)} \right) \right]
\frac{1}{1+\frac{e}{\hbar}
  {\bf{B}}\cdot{\bf{\Omega}}_{\bf k}^{\left(\pm\right)}}
\nonumber \\
& & \quad\times\left[v_x^{\left(\pm\right)}+\frac{e}{\hbar}B_x({\bf{v}}_{\bf k}^{\left(\pm\right)}\cdot{\bf{\Omega}}_{\bf k}^{\left(\pm\right)})\right] \nonumber \\
& & \quad\times\left[v_y^{\left(\pm\right)}+\frac{e}{\hbar}B_y({\bf{v}}_{\bf k}^{\left(\pm\right)}\cdot{\bf{\Omega}}_{\bf k}^{\left(\pm\right)})\right],
\label{eq:sxy}
\eea
where $f_{\rm{eq}}$ is the equilibrium Fermi-Dirac distribution function.
We compute the components of the positive energy state, denoted by superscript $(+)$
and the negative energy state, denoted by superscript $(-)$, separately.
Here, the magnetic field is given by $\mathbf{B}=(B_x,B_y,0)=(B\cos\phi,B\sin\phi,0)$.
In order to obtain the total magnetoconductivity, we take the sum of
$\sigma_{xx}^{(+)} + \sigma_{xx}^{(-)}$
and
$\sigma_{xy}^{(+)} + \sigma_{xy}^{(-)}$.
We also calculate the contribution from the other two Dirac cones.
The splitting of each Dirac cone in the $k_z$ direction
results in a twofold multiplication factor.

\begin{figure}[htbp]
  \includegraphics[width=0.4\textwidth]{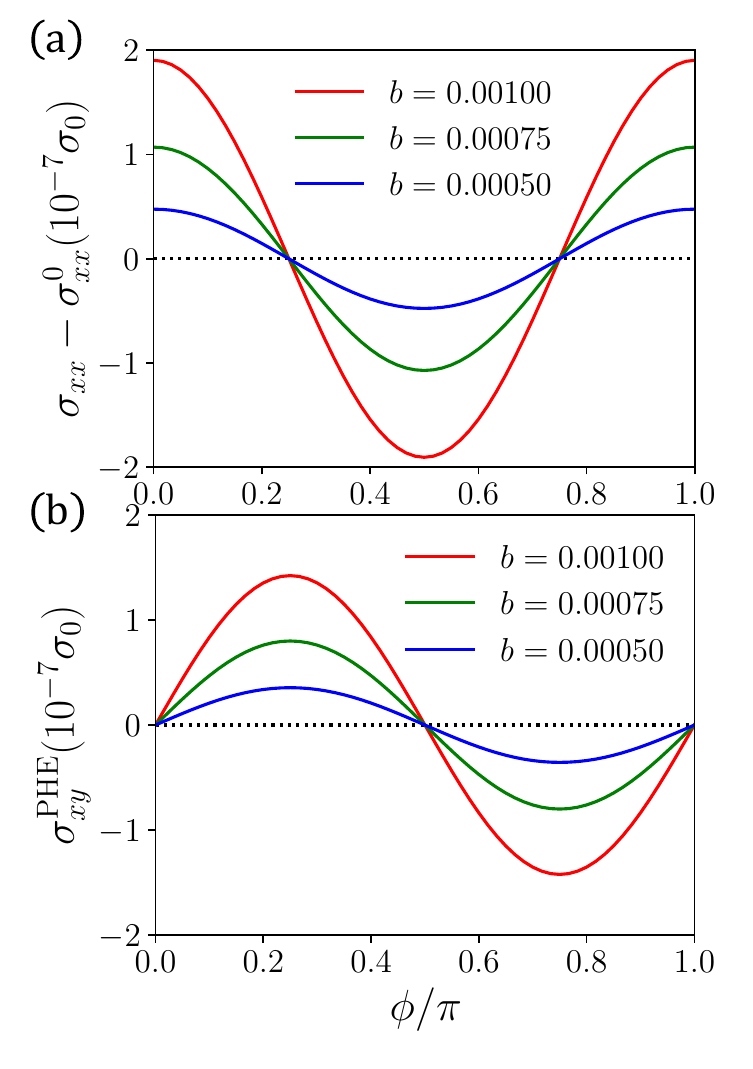}
  \caption{
    \label{fig:bDependence}
  (Color online)
  (a) Longitudinal conductivity and (b) planar Hall conductivity as the function of $\phi$
    for different values of $b$.
    The unit of conductivity is $\sigma_0=e^2\tau v/(2\pi^3 \hbar a c)$.
  The constant component is subtracted from $\sigma_{xx}$.
  The parameters are $\tilde{t_1}=0.10$, $\tilde{t_2}=0.05$, and $\eta=0.7$.
  For the Dirac point energies,
  we set $\varepsilon_{\rm D}/(\hbar v/a)=0.5$
  for two Dirac cones along the $k_z$ axis
  and $\varepsilon_{\rm D}/(\hbar v/a)=-0.4$ for the other two Dirac cones.
  }
\end{figure}

The result is shown in Fig.~\ref{fig:bDependence}.
%------------------------------------------------------------------------
We subtract the constant value $\sigma_{xx}^0$ from $\sigma_{xx}$,
and the oscillating component $\sigma_{xx}-\sigma_{xx}^0$
is shown in Fig.~\ref{fig:bDependence}(a).
As for $\sigma_{xy}$, we denote it as $\sigma_{xy}^{\rm PHE}$
in Fig.~\ref{fig:bDependence}(b)
to explicitly indicate that its contribution originates from the planar Hall effect.
They are plotted as the function of $\phi$
for different values of $b=\left(a/\ell_B \right)^2$ with $\ell_B=\sqrt{\hbar/eB}$
the magnetic length.
$b$ is defined as the dimensionless magnetic field parameter.
At $B=1$~T, $b=1.5\times 10^{-3}$.
The unit of conductivity is $\sigma_0=e^2\tau v/(2\pi^3 \hbar a c)$.
For the interlayer tunneling parameters, $\tilde{t}_1$ and $\tilde{t}_2$, we take
$\tilde{t}_1=0.10$ and $\tilde{t}_2=0.05$.
For the tilt parameter we take $\eta=0.7$.
This set of parameters is reasonable for \alphaI.
We note that both
$\sigma_{xx}-\sigma_{xx}^0$ and $\sigma_{xy}^{\rm PHE}$
exhibit the periodicity of $\pi$.
This oscillating behavior can be associated
with the chiral anomaly\cite{Nandy2017,Burkov2017}.
Qualitatively similar behavior is observed in a recent experiment\cite{Tajima2023a}.

%------------------------------------------------------------------------
%------------------------------------------------------------------------

\begin{figure}[htbp]
  \includegraphics[width=0.5\textwidth]{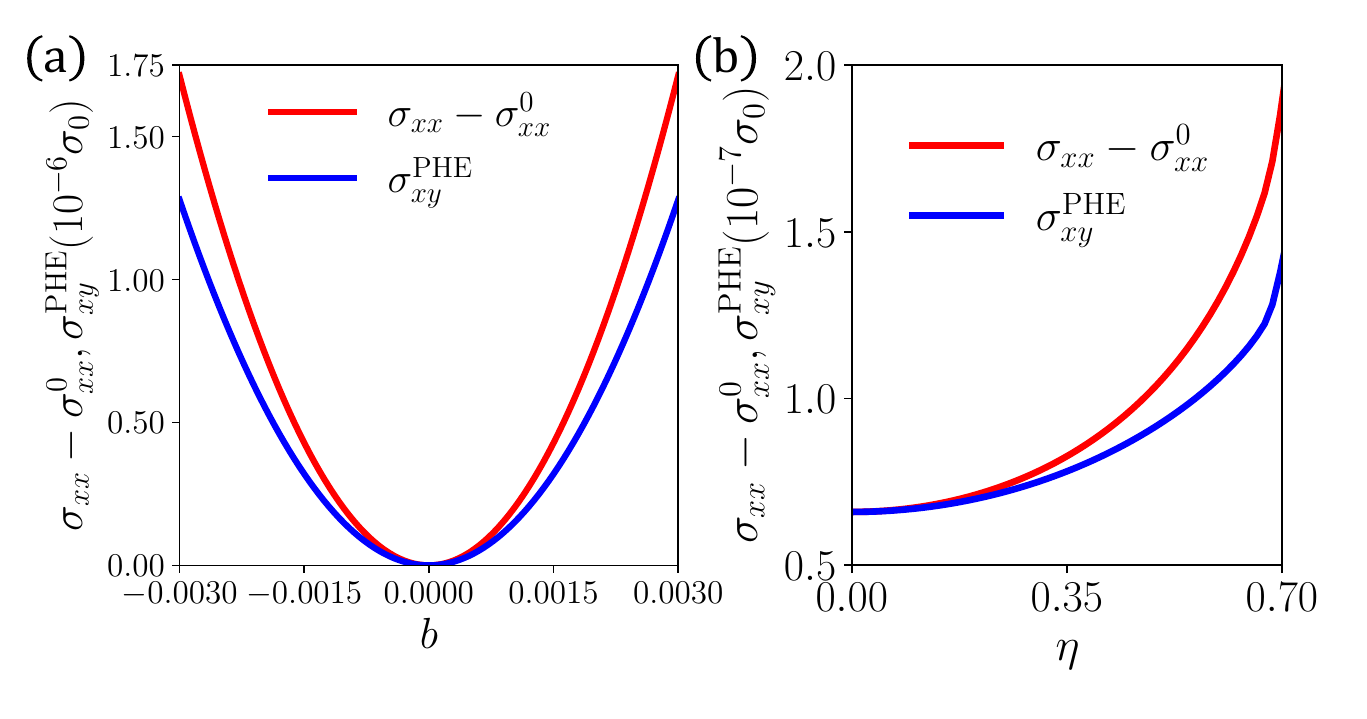}
  \caption{
    \label{fig:bDepAmp}
  (Color online)
    (a) The magnetic field dependence of the
    amplitudes of $\sigma_{xx}-\sigma_{xx}^0$
    and $\sigma_{xy}^{\rm PHE}$
    using the same parameters as in Fig.~\ref{fig:bDependence}.
    The amplitude varies quadratically with the magnetic field
    and there is no linear term.
    (b) The tilt parameter dependence
    of the amplitudes of 
    $\sigma_{xx}-\sigma_{xx}^0$
    and $\sigma_{xx}^{\rm PHE}$ at $b=0.001$.
  }
\end{figure}

We also examined the magnetic field parameter $b$ dependence
of the amplitude of
$\sigma_{xx}-\sigma_{xx}^0$ and $\sigma_{xy}^{\rm PHE}$
as shown in Fig.~\ref{fig:bDepAmp}(a).
We find that the amplitude varies quadratically with the magnetic field.
If there remains the effect associated with the tilt of the Dirac cone,
we may expect a linear dependence, but there is no such component.
This is understood by complete cancellation
between the contribution from the Dirac cones with opposite tilts
and chiralities.
We note that the amplitudes of
$\sigma_{xx}-\sigma_{xx}^0$
is slightly larger than $\sigma_{xy}^{\rm PHE}$.
This behavior is qualitatively in agreement with experimental observations,
where the amplitude of
$\sigma_{xx}-\sigma_{xx}^0$
is ten times larger than that of $\sigma_{xy}^{\rm PHE}$ at 3~T\cite{Tajima2023a}.
The difference of the amplitutdes is associated with the inteplay between
the tilt parameter dependence of the group velocity
and the density of states.
To make clear the tilt parameter dependence,
we calculate $\eta$ dependence of 
$\sigma_{xx}-\sigma_{xx}^0$
and $\sigma_{xy}^{\rm PHE}$
as shown in Fig.~\ref{fig:bDepAmp}(b).
When $\eta=0$, there is no difference in the amplitudes
of $\sigma_{xx}-\sigma_{xx}^0$
and $\sigma_{xy}^{\rm PHE}$.
Their difference increases as we increase $\eta$.
However, the result depends on the choice of two values of $\varepsilon_{\rm D}$.
If we take a different set of values for $\varepsilon_{\rm D}$,
we obtain a different $\eta$ dependence.
The energy dispersion exhibits particle-hole symmetry;
however, the integrands in Eqs.~(\ref{eq:sxx}) and (\ref{eq:sxy}) do not.
Consequently, the $\eta$ dependence of 
$\sigma_{xx}-\sigma_{xx}^0$
and $\sigma_{xy}^{\rm PHE}$ is non-trivial.

%------------------------------------------------------------------------
%------------------------------------------------------------------------

To conclude, we have shown that the magnetoconductivity exhibit PHE
in a realistic model for \alphaI.
Since \alphaI does not show any indication of ferromagnetism\cite{Konoike2022},
the presence of PHE suggests the chiral anomaly effect
that is associated with a three-dimensional Dirac semimetal.
While our analysis is confined to a small magnetic field range
due to the utilization of the semiclassical Boltzmann equation,
we anticipate the occurrence of the PHE at high magnetic fields,
provided there is no qualitative change between the low and high magnetic field
regimes.
This seems to be consistent with the recent experiment\cite{Tajima2023a}.
In conjunction with the experimental findings\cite{Tajima2023,Tajima2023a},
our results provide strong support for the classification of \alphaI
as a three-dimensional Dirac semimetal under conditions of low temperatures and high pressures.

\begin{acknowledgments}
  We thank N. Tajima for helpful discussions and sharing experimental data.
  The research was supported by JSPS KAKENHI Grant Number 22K03533.
\end{acknowledgments}

\bibliography{../../../refs/lib}
\end{document}